\documentclass[fleqn,usenatbib]{mnras}

\usepackage{newtxtext,newtxmath}

\usepackage[T1]{fontenc}

\DeclareRobustCommand{\VAN}[3]{#2}
\let\VANthebibliography\thebibliography
\def\thebibliography{\DeclareRobustCommand{\VAN}[3]{##3}\VANthebibliography}

\usepackage{graphicx}	
\usepackage{amsmath}	

\newcommand\maps{\ref@jnl{M\&PS}}

\title[Icy dead zones around low-mass stars]{Formation of super-Earths in icy dead zones around low-mass stars }

\author[D. Vallet et al.]{
David Vallet,$^{1}$
Anna C. Childs,$^{2,3,4}$
Rebecca G. Martin,\thanks{E-mail: rebecca.martin@unlv.edu}$^{2,3}$
Mario Livio$^{2,3}$ and Stephen Lepp$^{2,3}$
\\
$^1$Department of Mechanical Engineering, University of Nevada, Las Vegas, 4505 South Maryland Parkway, Las Vegas, NV 89154, USA \\
$^2$Nevada Center for Astrophysics, University of Nevada, Las Vegas,
4505 South Maryland Parkway, Las Vegas, NV 89154, USA\\
$^3$Department of Physics and Astronomy, University of Nevada, Las Vegas, 4505 South Maryland Parkway, Las Vegas, NV 89154, USA\\
$^4$Center for Interdisciplinary Exploration and Research in Astrophysics (CIERA) and Department of Physics and Astronomy, \\
Northwestern University, 1800 Sherman Ave, Evanston, IL, 60201, USA \\
}

\date{Accepted XXX. Received YYY; in original form ZZZ}

\pubyear{2022}

\begin{document}
\label{firstpage}
\pagerange{\pageref{firstpage}--\pageref{lastpage}}
\maketitle

\begin{abstract}
While giant planet occurrence rates increase with stellar mass, occurrence rates of close-in super-Earths decrease. This is in contradiction to the expectation that the total mass of the planets in  a system scale with the protoplanetary disc mass and hence the stellar mass.  Since the snow line plays an important role in the planet formation process we examine differences in the temperature structure of protoplanetary gas discs around stars of different mass. Protoplanetary discs likely contain a dead zone at the midplane that is  sufficiently cold and dense for the magneto-rotational instability to be suppressed. As material builds up, the outer parts of the dead zone may be heated by self-gravity. The temperature in the disc can be below the snow line temperature far from the star and in the inner parts of a dead zone.
The inner icy region has a larger radial extent around smaller mass stars. The increased mass of solid icy material may allow for the in situ formation of larger and more numerous planets close to a low-mass star. Super-Earths that form in the inner icy region  may have a composition that includes a significant fraction of volatiles. 
\end{abstract}




\section{Introduction}

Explaining differences in planetary system architectures around stars with different masses can help us to understand the planet formation process more generally.  The probability of hosting a giant planet increases with stellar mass \citep[e.g.][]{Johnson2007,Johnson2010}. Giant planets are found around less than 10\% of G type stars and this fraction decreases with decreasing stellar mass. Giant planets with a semi major axis in the range of $3-7\,\rm au$ are  around 6.2\% of G type stars \citep{Wittenmyer2016,Wittenmyer2020}.
The upper limit on the fraction of M-dwarfs with hot Jupiters is 0.34\% \citep{Obermeier2016} while for FGK stars it's 1\% \citep{Johnson2007,Beleznay2022}. There may be a slight decrease in hot Jupiter occurrence rate with stellar mass for higher mass GFA stars \citep{Zhou2019}.     On the other hand, the occurrence rate of super-Earths with orbital period $<100\,\rm days$ is more than three times higher around M-dwarfs compared to F stars \citep{Howard2012,Dressing2013,Mulders2015, Mulders2015c,Mulders_2018,Mulders_2021,Sabotta2021}. The average mass of heavy-elements in planets is inversely proportional to stellar mass with F stars having around $4\,\rm M_\oplus$ and M-dwarfs having around $7\,\rm M_\oplus$ \citep{Mulders2015c}.  M-dwarfs have a higher fraction of stars with planetary systems and a higher planetary multiplicity \citep{Yang2021,He2021}. However, this seems to be in contradiction to the expectation from the protoplanetary disc masses \citep[e.g.][]{Schlecker2022}. More massive stars have more massive protoplanetary gas discs and therefore more material available for planet formation \citep[e.g.][]{Mordasini2012}.

There are a number of suggested mechanisms for forming close-in super-Earths although most require formation farther out followed by inward migration \citep{McNeil2010,Haghighipour2013,Inamdar2015,Schlichting2018,Zawadzki2021}. One suggested explanation for the excess of super-Earths around low-mass stars is that pebble accretion in the inner regions is shut off when a giant planet forms and opens a gap in the disc \citep{Lambrechts2019,Mulders_2021}. This suppresses the growth of close-in planets.  However, some studies suggest that super-Earths may be present even with cold giant planets \citep{Barbato2018,Zhu2018,Bryan2019}. 
Fragmentation of a massive disc around an M-dwarf may be possible, but it requires a disc with a mass of around 30\% of the star \citep{Boss2006,Backus2016,Mercer2020,Haworth2020}.

The snow line radius in a protoplanetary disc plays an important role in the planet formation process. It denotes the distance from the star outside of which water is found in a solid form and it occurs at temperatures of around $T_{\rm snow}=170\,\rm K$ \citep{Hayashi1981,Podolak2004,Lecaretal2006}.
Giant planets are thought to form in a protoplanetary disc at temperatures below the snow line temperature. The solid material available for planet formation is about 4 times higher below the snow line temperature \citep{Pollack1996,Kennedy2008}. The locations of the observed giant planets indeed peaks around the snow line radius around all stellar types \citep{Fernandes_2019,Childs2022}. 

If the disc around a solar-mass star is fully turbulent, the snow line radius can move inwards of $1\,\rm au$ within the disc lifetime \citep[e.g.][]{Ida2005,Davis2005,GaraudandLin2007,Min2011}. However, the compositions of asteroids in the asteroid belt in the solar system suggest that the snow line radius was around $2.7\,\rm au$ in the solar system at the time of disc dissipation \citep[e.g.][]{DeMeo2014}. It has previously been suggested that a dead zone that becomes self-gravitating can provide the additional heating necessary to keep the snow line farther out \citep{Martin2012}. A dead zone is a region of the disc in which the magneto-rotational instability \citep[MRI,][]{BH1991} is suppressed due to the low temperature and high density of the disc \citep[e.g.][]{Gammie1996,Armitage2001,Fleming2003,Zhuetal2010a,MartinandLubow2011,Bai2011,Delage2022}. The inner parts of the dead zone can  add an additional icy region to the disc structure \citep{Martin2013snow,Xiao2017}. In addition to the differences in the temperature structure, the dead zone leads to a pile up of material and a larger surface density than in the fully turbulent disc model \citep{Martin2016se}. 

In this Letter we explore differences in the structure of protoplanetary discs around stars of different masses. In  Section~\ref{steady} we consider fully turbulent steady state disc solutions and compare those to  a disc with a dead zone. We show that the inner icy region in a dead zone is larger around an M-dwarf than around a solar-mass star, and we suggest that this facilitates the formation of super-Earths around low-mass stars.  We present our conclusions in Section~\ref{conc}.

\section{Disc structure}
\label{steady}

 The gas in an accretion disc orbits the central star of mass $M$  at orbital radius $R$ with Keplerian frequency given by $\Omega = \sqrt{G M/R^3}$ \citep{LP1974,Pringle1981}. We explore disc structures with a steady accretion rate through the disc so that
\begin{equation}
\dot M =3 \pi \nu \Sigma,
\label{mdot}
\end{equation}
where $\Sigma$ is the surface density and the viscosity is given by
\begin{equation}
    \nu= \alpha \frac{c_{\rm s}^2}{\Omega},
\end{equation}
where $\alpha$ is the \cite{SS1973} viscosity parameter. The sound speed is
\begin{equation}
c_{\rm s}=\sqrt{\frac{{\cal R}T}{\mu}}, 
\end{equation}
where $T$ is the disc temperature, $\cal R$ is the gas constant and $\mu=2.3$ is the mean molecular weight.
The mid-plane temperature of the disc is found by solving
\begin{equation}
    \sigma T^4 =\frac{9}{8} \Omega^2 \tau \nu \Sigma,
    \label{temps}
\end{equation}
where the optical depth is given by
\begin{equation}
    \tau=\frac{3}{8}\kappa \frac{\Sigma}{2}.
\end{equation}
For simplicity, we take the opacity to be $\kappa=0.02\,T^{4/5}\,\rm cm^2 g^{-1}$ \citep{Armitage2001}. The opacity at the snow line increases as a result of increased solid abundance \citep{Bell1994}. We note that the temperature and therefore the snow line radius is not very sensitive to the opacity. In order to find steady disc solutions with $\dot M=\,$const, we solve equations~(\ref{mdot}) and~(\ref{temps}) for $\Sigma$ and $T$.

\subsection{Fully turbulent disc structure}
\label{sec:turb}

We first consider the conditions required for a fully turbulent disc solution to exist at all radii in the disc. A dead zone forms at the disc midplane where the temperature is below the critical value, $T<T_{\rm crit}$, and the surface density is above the critical, $\Sigma>\Sigma_{\rm crit}$. The critical temperature is around $T_{\rm crit}=800\,\rm K$ \citep{Gammie1996} while the critical surface density is less well constrained. This is the surface density in the surface layers of the disc that is ionised by external sources such as cosmic rays or X-rays from the star \citep{Glassgoldetal2004}. A disc is fully turbulent if the surface density at the radius where $T=T_{\rm crit}$ is less than $\Sigma_{\rm crit}$. This surface density is shown in Fig.~1 in \cite{Martin2013snow} and we note that this is independent of the stellar mass. 
If the disc is fully turbulent, it transitions directly from a thermally ionised inner region to an externally ionised outer region. 
If $\Sigma_{\rm crit}\lesssim 20\,\rm g\, cm^{-2}$, then the disc contains a dead zone throughout its lifetime for all stellar masses. 

The active layer surface density may be different around M-dwarfs compared to solar-mass stars depending on the dominant ionisation source - it may be similar if the main source is cosmic rays but different if X-rays from the central star dominate. The lower luminosity of a lower mass star leads to a lower X-ray ionisation rate and therefore a smaller critical surface density and a larger dead zone \citep{Delage2022}.

The structure of a fully turbulent disc is found by solving equations~(\ref{mdot}) and~(\ref{temps}) with a constant $\alpha=0.01$. The green lines in Fig.~\ref{plot2} show the radius at which $T=T_{\rm crit}=800\,\rm K$ in this model. For smaller radii, there is a fully turbulent steady state solution with $T>T_{\rm crit}$. However for larger radii there is no solution with a sufficiently high temperature to avoid a dead zone forming.

\subsection{Stellar irradiation}
\label{sec:irr}

Close to the star, the irradiation from the star may dominate the temperature of the disc over viscous heating.  This provides a lower limit to the temperature at a given radius.   The irradiation temperature is
\begin{equation}
    T_{\rm irr}=\left(\frac{\alpha_{\rm irr}}{2}\right)^{1/4}\left(\frac{R_\star}{R}\right)^{1/2}T_\star
\end{equation}
\citep{Chiang1997}, where $T_\star$ is the temperature of the star and $R_\star$ is its radius. For a flat disc geometry with a constant disc aspect ratio, $H/R=\,$const,
\begin{equation}
  \alpha_{\rm irr}=0.4\left(\frac{R_\star}{R}  \right)
\end{equation}
 and for a flared disc in hydrostatic equilibrium with $H/R\propto R^{1/2}$, 
 \begin{equation}
  \alpha_{\rm irr}=0.005\left( \frac{R}{\rm au}\right)^{-1}+0.05\left(\frac{R}{\rm au}\right)^{2/7}     
 \end{equation}
\citep{Kenyon1987,Chiang1997}.

\begin{figure*}
\begin{center}
\includegraphics[width=\columnwidth]{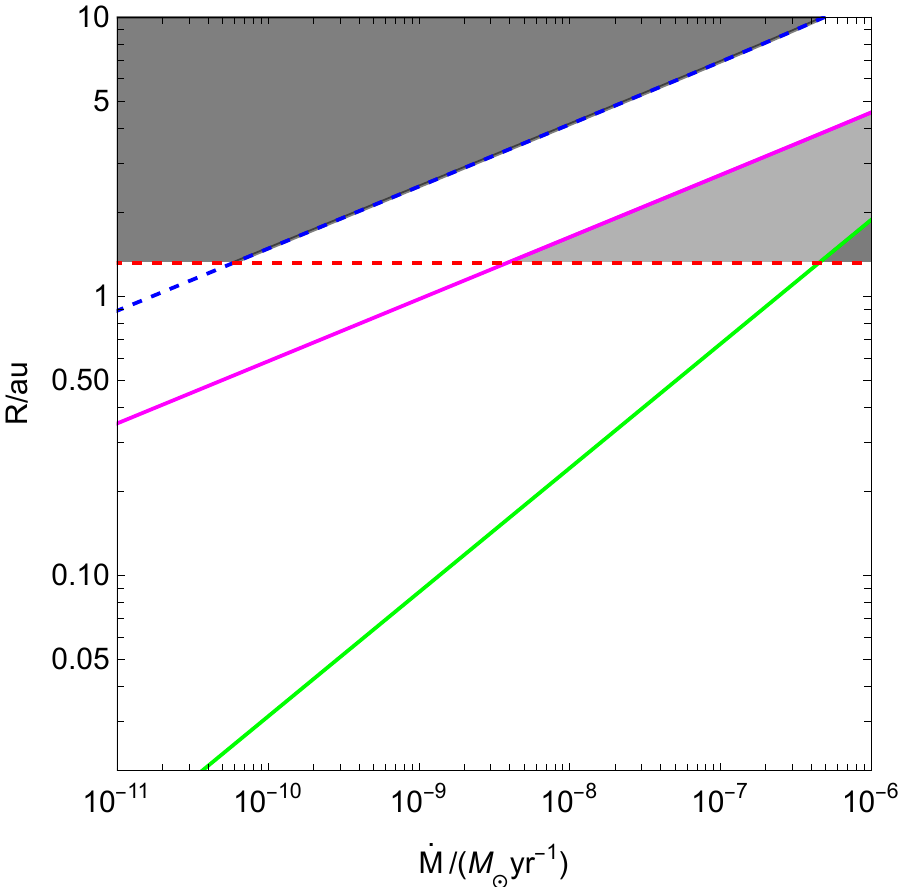}
\includegraphics[width=\columnwidth]{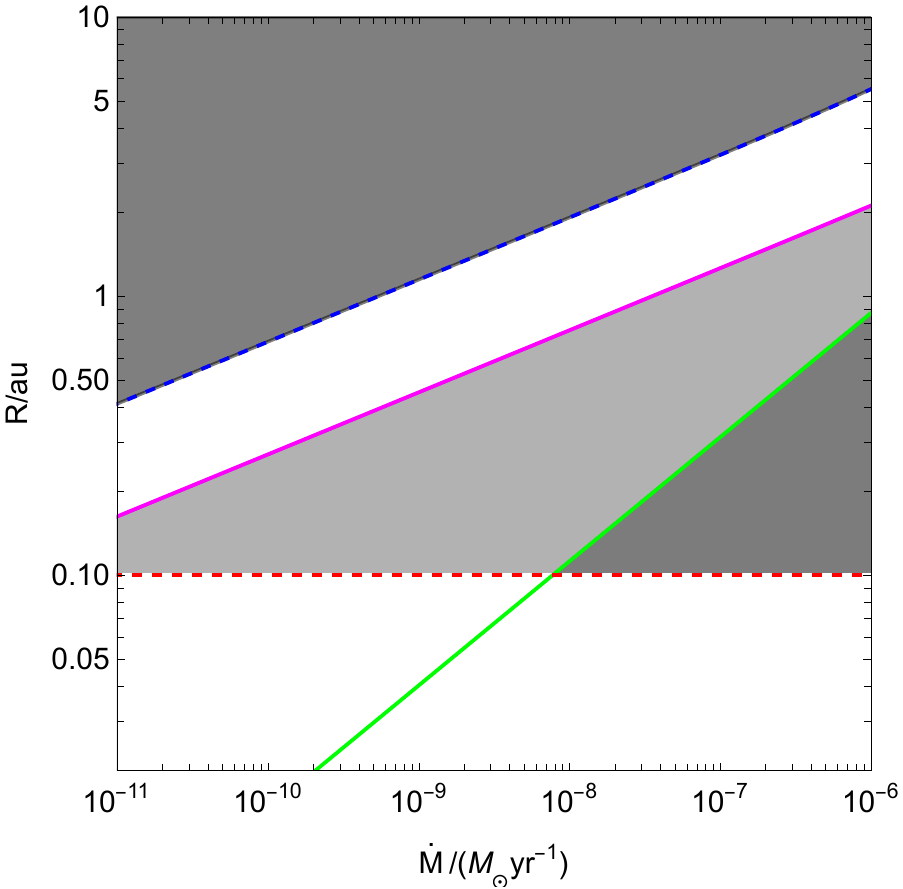}
	\end{center}
    \caption{Shaded regions show the icy regions of a disc with a dead zone for different steady state accretion rate around a star with mass $M=1\,\rm M_\odot$ (left) and $M=0.1\,\rm M_\odot$ (right). The green solid line shows where the fully turbulent disc model has $T=T_{\rm crit}=800\,\rm K$ (see Section~\ref{sec:turb}). The red dashed lines show where the irradiation temperature equals the snow line temperature for the flared disc model (see Section~\ref{sec:irr}). The blue dashed lines show the snow line radius in a self-gravitating disc and the magenta lines show where  $T=T_{\rm crit}$ (see Section~\ref{GM}). Between the magenta and green lines there are no steady solutions.  The outer extent of the inner icy region can vary in time between the magenta and green lines. Since the dead zone region is not in a steady state, the accretion rate on to the star is lower than the steady rate shown.
    }
    \label{plot2}
\end{figure*}

We find the snow line radius in an irradiated disc by setting $T_{\rm irr}=T_{\rm snow}=170\,\rm K$.
For a pre-main-sequence solar-mass star we take $T_\star=4000\,\rm K$ and $R_{\star}=3\,\rm R_\odot$ and for an M-dwarf with mass $M=0.1\,\rm M_\odot$ we take $T_\star=2850$ and $R_\star=0.4\,\rm R_\odot$ \citep[e.g.][]{Hasegawa2010}.
For a flat disc geometry,  $R_{\rm snow}=0.56\,\rm au$ for solar mass star and $R_{\rm snow}=0.047\,\rm au$ for the M-dwarf.  These values are the minimum possible values for the snow line radius. For the flared disc model,  $R_{\rm snow}=1.31\,\rm au$ for the solar mass star and $R_{\rm snow}=0.10\,\rm au$ for the M-dwarf. These radii are independent of the accretion rate through the disc and are shown in Fig.~\ref{plot2} as the red dashed lines.

\subsection{Self-gravitating steady state disc structure}
\label{GM}

We now consider a situation in which a dead zone forms in the disc. This requires $\Sigma_{\rm crit}$ to be sufficiently low, so that there is not a steady state solution throughout the disc. In this case, material builds up in the dead zone until the outer parts become self-gravitating. We  assume that $\Sigma \gg \Sigma_{\rm crit}$ and we ignore the thin active layer.
A disc becomes self gravitating when the \cite{Toomre1964} parameter, $Q=c_{\rm s}\Omega/\pi G \Sigma$, becomes less than its critical value of $Q_{\rm crit}=2$. In this case, the viscosity in the disc can be approximated with 
\begin{equation}
    \alpha=\alpha_{\rm c}\left(
    \left(\frac{Q_{\rm crit}}{Q}\right)^2-1
    \right)
\end{equation}
\citep[e.g.][]{LP1987,LP1990}, where we take $\alpha_{\rm c}=0.01$. We can find steady disc solutions with this parameter $\alpha$ by again solving  equations~(\ref{mdot}) and~(\ref{temps}).

The blue dashed lines in Fig.~\ref{plot2} show the snow line radius in a disc with a self-gravitating dead zone.  The snow line in this disc is much farther out than the snow line in the fully turbulent disc, and in better agreement with that observed in the solar system \citep[see also][]{Martin2012,Martin2013snow}.
The magenta lines in  Fig.~\ref{plot2} show where this steady solution has $T=T_{\rm crit}$. This is the minimum radius for this steady solution since if the temperature reaches $T_{\rm crit}$, then the MRI is triggered.

\subsection{Icy region in the dead zone}

A disc with a dead zone is not a steady disc at all radii. At the radius where the dead zone becomes self-gravitating, there is a local peak in the temperature and the surface density \citep[see e.g. Fig.~2 in][]{MartinandLubow2013prop}. As material builds up, the peak moves inwards and increases until the critical temperature for the MRI to be triggered is reached.  This causes an accretion outburst when the dead zone becomes fully MRI active for a short period of time.

 In Fig.~\ref{plot2} we shade the regions of the disc in which the temperature can be below the snow line temperature for a solar-mass star (left) and an M-dwarf (right).  There may be an inner icy region within the dead zone in the disc, where the temperature is below that of the snow line temperature. 
The outer edge of this region is bounded by the location where the disc becomes self-gravitating. This varies in time in the region in which there is no steady state disc solution for a given accretion rate \citep[see Fig. 2 in][]{MartinandLubow2013prop}. The unstable region is bounded by where the fully turbulent solution has a temperature $T_{\rm crit}$ (the green line) and where the self-gravitating steady solution has  a temperature $T_{\rm crit}$ (the magenta line).  The radius at which the stellar irradiation temperature equals the snow line temperature provides an inner limit to this region (see the red dashed lines). 
Note that the shaded region is the maximum radial extent of the inner icy region in the limit that $\Sigma_{\rm crit}$ is small. If $\Sigma_{\rm crit}$ is sufficiently large, then the disc can transition to a fully MRI active disc for low accretion rates \citep[see Section~\ref{sec:turb} and][]{MartinandLubow2013dza}. 

The accretion rate in these figures is the steady state accretion rate. Since the inner parts of the disc are not in a steady state, this does not correspond to the accretion rate on to the star. Material is building up within the dead zone region meaning that the accretion rate on to the star may be much smaller than the steady rate. 

The extent of the inner icy region can be much larger around an M-dwarf than around a solar-mass star. We therefore suggest that this inner icy region, that increases in size with decreasing stellar mass, can help to explain the observed trend of increasing number of super-Earth mass planets around low mass stars. Note that the inner icy region in inherently unstable. It exists in the time between outbursts. During the outburst, ice in the inner regions of the disc may be evaporated and then it recondenses \citep[e.g.][]{Hubbard2017}. However, the time between outbursts may be much longer around an M-dwarf than around a solar-mass star since the infall accretion rate is considerably lower, so that it takes longer to build up the material required for an outburst.

There is an observed trend that the metallicity of a star increases with its mass \citep[e.g.][]{Ghezzi2010}. While the metallicity of a star is positively correlated with the presence of giant planets \citep{Fischer2005}, there appears to be no correlation with lower mass planets \citep{Kutra2021}. We suggest that this is because differences in the amount of solid material in the inner parts of a disc are dominated by the temperature of the disc rather than the initial ratio of dust-to-gas.

\section{Discussion and Conclusions}
\label{conc}

The occurrence rate of super-Earth planets increases with decreasing stellar mass. This is in contradiction to what is predicted by a fully turbulent disc model in which the disc mass scales with the stellar mass. However, a disc model with a dead zone can contain an additional inner icy region. We have shown that the radial extent of this inner icy region increases with decreasing stellar mass. The inner icy region can contain more solid icy mass than predicted by a fully turbulent disc model.  While the disc around a solar mass star is more massive than a disc around an M-dwarf, there may be more solid material around the M-dwarf because of the low temperature. 

This icy region may allow for the rapid formation of super-Earths through core accretion \citep[see also][]{Kennedy2006,kennedy2007}. Pebble accretion is another theory for planet formation in which centimeter sized pebbles accrete directly onto forming protoplanets in a gasous disc \citep{Ormel2010}. Ices are stickier than silicate dust and the fragmentation velocity may be 10 times smaller in regions with a temperature above that at the snow line \citep[e.g.][]{Mulders_2021} and therefore pebble sizes can be much smaller \citep[e.g.][]{Levison2015,Morbidelli2015}. The inner icy region in a disc model with a dead zone would allow larger pebble sizes and therefore larger mass planets to form.  

We conclude that models of planet formation around low-mass stars, whether by core accretion or pebble accretion, need to include the inner icy region of a dead zone and the associated increase in solid icy material there.  The temperature of the disc in the planet forming region has implications for the composition of the forming planets  \citep[e.g.][]{Alibert2017}. Our gas disc models suggest that super-Earths around low-mass stars may form in situ with significant amounts of water and volatiles. A small amount of inward migration may still be required to explain some observed systems. The TRAPPIST~1 planets have semi-major axes in the range $0.01-0.06\,\rm au$ while the icy dead zone would have been at distances $\gtrsim 0.1\,\rm au$. 

There is a bimodal radius distribution of small and close-in exoplanets that is thought to be a result of atmospheric loss either through photoevaporation by the high-energy radiation from the star \citep[e.g.][]{Owen2013,Owen2017} or core-powered mass loss \citep{Gupta2019,Gupta2020}. These planet evolutionary models suggest that the composition of super-Earths is rocky, rather than icy. However, more recently it has been suggested that water worlds might be present around later type stars since the radius gap is a density gap \citep{Luque2022}. This would be in agreement with the model presented in this work in which the super-Earths around a low mass star form in the inner icy region. Alternatively,  formation might start outside the snowline and then proceed inside after some migration \cite[e.g][]{Ormel2017}.

The temperature of the disc determines the distribution of solids. In regions of the disc where the temperature is set by the absorption of reprocessed stellar light by the solids the snow line radius can be thermally unstable \citep{Owen2020}.  This occurs  for moderate optical depths, where heating by absorption of reprocessed stellar light from the disc’s atmosphere is optically thick but cooling is optically thin. When the snow line is unstable, its location oscillates in time. If the inner snow line discussed in this work is thermally unstable this could lead to a larger inner icy region during part of the evolution. This should be investigated in future work with more detailed models.

\section*{Acknowledgements}
We thank an anonymous referee for useful comments that improved the manuscript. We acknowledge support from NASA through grant 80NSSC21K0395.  AC acknowledges support from the NSF through grant NSF AST-2107738.


\section*{Data Availability}

 The data underlying this article will be shared on reasonable request to the corresponding author.
 



\bibliographystyle{mnras}








\bsp	
\label{lastpage}
\end{document}